\documentclass[prl,aps,twocolumn,superscriptaddress]{revtex4-1}

\usepackage{amsmath,amssymb,amsfonts,color,graphicx,tabularx}

\usepackage[unicode=true,colorlinks=true]{hyperref}

\hypersetup{linkcolor=blue,citecolor=blue,urlcolor=blue}

\begin{document}

\title{Exactly solvable quantum impurity model with inverse-square interactions}

\author{Hong-Hao Tu}
\email{hong-hao.tu@tu-dresden.de}
\affiliation{Institut f\"ur Theoretische Physik, Technische Universit\"at Dresden, 01062 Dresden, Germany}

\author{Ying-Hai Wu}
\email{yinghaiwu88@hust.edu.cn}
\affiliation{School of Physics and Wuhan National High Magnetic Field Center, Huazhong University of Science and Technology, Wuhan 430074, China}

\begin{abstract}
We construct an exactly solvable quantum impurity model which consists of spin-1/2 conduction fermions and the spin-1/2 magnetic moment. The ground state is a Gutzwiller projected Fermi sea with non-orthonormal modes and its wave function in the site-occupation basis is a Jastrow-type homogeneous polynomial. The parent Hamiltonian has all-to-all inverse-square hopping terms between the conduction fermions and inverse-square spin-exchange terms between the conduction fermions and the magnetic moments. The low-lying energy levels, spin-spin correlation function, and von Neumann entanglement entropy of our model demonstrate that it exhibits the essential aspects of spin-1/2 Kondo physics. The machinery developed in this work can generate many other exactly solvable quantum impurity models.
\end{abstract}

\maketitle

\date{\today}

{\em Introduction} --- The interplay between the localized degree of freedom and itinerant electrons has been a central subject of condensed matter physics in the past decades~\cite{hewson1997,coleman2015}. The localized degree of freedom is usually called impurity and the simplest example is spin-1/2 magnetic moments due to unpaired $d$ or $f$ electrons in solids. The unpaired electron is screened by the itinerant electrons via spin-exchange interactions at low temperatures and a spin-singlet state emerges. Quantum impurity physics has also been investigated in quantum dots~\cite{gordon1998,cronenwett1998}, magnetic adatoms on metallic surfaces~\cite{madhavan1998}, carbon nanotubes~\cite{nygard2000,herrero2005}, and quantum Hall devices~\cite{iftikhar2015,keller2015,iftikhar2018}. The impurity may even be realized in a non-local manner using Majorana zero modes in topological superconductors~\cite{beri2012,crampe2013,tsvelik2013,altland2014,eriksson2014,kashuba2015}.

The theoretical understanding of quantum impurity physics begins with two models proposed by Anderson and Kondo, which are related by the Schrieffer-Wolff transformation~\cite{anderson1961,kondo1964,schrieffer1966}. To explain a mysterious change of resistance in certain metallic samples with dilute magnetic atoms, Kondo constructed a model with one magnetic moment coupled to a bath of conduction electrons via a spin-exchange term. The simplicity of this model is highly deceptive as manifested by divergences in straightforward perturbative calculations, which calls for drastically different approaches. Based on the idea of renormalization, Wilson developed the numerical renormalization group (NRG) method and brought out a rather complete solution of the Kondo problem~\cite{wilson1975,krishna1980a,bulla2008}. The Kondo problem elucidates the concept of asymptotic freedom and serves as a vivid demonstration of correlation effects in many-body systems.

Further insights into the Kondo physics were gathered along several directions. The NRG results helped Nozi\`eres to formulate a Fermi liquid theory for the low-energy physics of the spin-1/2 Kondo model~\cite{nozieres1974}. The magnetic moment is only coupled to one particular mode of the conduction fermions and other modes in the conduction bath remain free. Affleck and Ludwig applied boundary conformal field theory (CFT) to study Kondo physics~\cite{affleck1990a,affleck1991c}. This approach reconstructs the Fermi liquid picture for the spin-1/2 Kondo model, but its utility extends to more complicated quantum impurity systems that are not amenable to the Fermi liquid description. Andrei and Wiegmann independently constructed a Kondo model which can be solved exactly using the Bethe ansatz~\cite{andrei1980,wiegmann1980,andrei1983,tsvelik1983}. It is defined on a semi-infinite continuous chain populated by fermions with linear dispersion relation. The Kondo physics has also been explored in some other exactly solvable models~\cite{andrei1984,schlottmann1991,sorensen1993,frahm1997,wangyp1997}.

In this Letter, we propose an exactly solvable quantum impurity model in a one-dimensional open chain with spin-1/2 conduction fermions and one spin-1/2 magnetic moment at one end of the chain. This model is very different from the one by Andrei and Wiegmann: it can be defined on any finite-size lattices; the conduction fermions have a quadratic dispersion relation in a special basis; the lowest eigenstates in all cases with an odd number of conduction fermions are known. The low-lying energy levels, entanglement entropy, and spin-spin correlation function of our model clearly demonstrate that it exhibits spin-1/2 Kondo physics. It is further suggested that our model may be non-integrable based on the level statistics of the full energy spectrum.

The most appealing property of our model is the existence of highly structured eigenstates that can be expressed in three different but equivalent ways: Gutzwiller projected Fermi seas with non-orthogonal basis states, Jastrow-type homogeneous polynomial wave functions, correlators of the CFT for the system. The Gutzwiller approach elucidates Nozi\`eres' Fermi liquid picture in an analytically tractable manner. The CFT approach is reminiscent of the well-established connection between edge CFT and bulk wave function in quantum Hall systems~\cite{moore1991}. The machinery developed in this work opens up the exciting possibility of constructing a variety of exactly solvable quantum impurity models.

\begin{figure}
\includegraphics[width=0.30\textwidth]{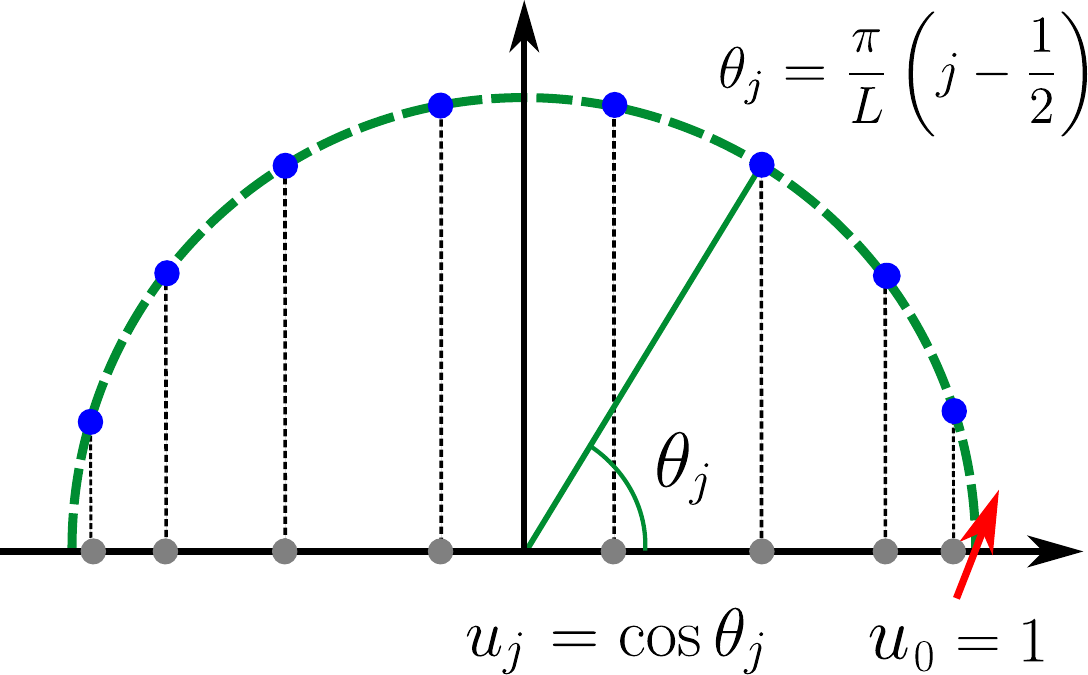}
\caption{Schematics of the quantum impurity model. The red arrow represents a spin-1/2 magnetic moment. The blue dots are lattice sites occupied by conduction fermions. The angular position of the $j$-th site is $\theta_{j}=\tfrac{\pi}{L}(j-1/2)$ (except for $\theta_0=0$) and its projection on the horizontal axis is $u_{j}=\cos\theta_{j}$.}
\label{Figure1}
\end{figure}

{\em Wave function} --- The system of our interest is an open chain with $L+1$ sites ($L$ can be even or odd) labeled by $j=0,1,\cdots,L$ [Fig.~\ref{Figure1}(a)]. It is placed on the semi-circle with unity radius and projected onto the real line $[-1,1]$. The angular position of the $j$-th site is $\theta_{j}=\tfrac{\pi}{L}(j-1/2)$ for $1{\leq}j{\leq}L$ and $\theta_{j}=0$ for $j=0$, and the associated linear position is $u_{j}=\cos\theta_{j}$. The motivation for this choice of coordinate will become clear later. The $1{\leq}j{\leq}L$ sites are populated by spin-1/2 conduction fermions described by creation (annihilation) operators $c^{\dag}_{j,\sigma}$ ($c_{j,\sigma}$) with $\sigma=\uparrow,\downarrow$ being the spin index. The $j=0$ site is occupied by a spin-1/2 magnetic moment described by operators ${\mathbf S}_{0}$. The magnetic moment can be represented using Abrikosov fermions as ${\mathbf S}_{0}=\frac12\sum_{\sigma\sigma'} c^{\dag}_{0,\sigma}{\vec\tau}_{\sigma\sigma'}c_{0,\sigma'}$ with $\vec{\tau}$ being the Pauli matrices~\cite{coleman2015}. The redundance caused by this representation is removed by imposing the single-occupancy constraint $\sum_{\sigma}c^{\dag}_{0,\sigma}c_{0,\sigma}=1$.

The many-body state to be investigated is
\begin{eqnarray}
|\Psi\rangle &=& \sum_{\{n^\uparrow_j\},\{n^\downarrow_j\}} \Psi(\{n^\uparrow_j\},\{n^\downarrow_j\}) \,
             (c^{\dag}_{0,\uparrow})^{n^\uparrow_0} \cdots (c^{\dag}_{L,\uparrow})^{n^\uparrow_{L}} \nonumber \\
   &\phantom{=}& \times (c^{\dag}_{0,\downarrow})^{n^\downarrow_0} \cdots (c^{\dag}_{L,\downarrow})^{n^\downarrow_{L}} |0\rangle .
\label{eq:wavefunc1}
\end{eqnarray}
The coefficient
\begin{eqnarray}
\Psi(\{n^\uparrow_j\},\{n^\downarrow_j\}) &=& \delta_{n_{0}=1} \delta_{\sum_{j}n^{\uparrow}_{j}=\sum_{j}n^{\downarrow}_{j}=M} \nonumber \\
&\phantom{=}& \times \prod_{\sigma=\uparrow,\downarrow} \prod_{0{\leq}j<k{\leq}L} (u_{j}-u_{k})^{n^\sigma_{j} n^\sigma_{k}}
\label{eq:wavefunc2}
\end{eqnarray}
is the wave function in the site-occupation basis, where $n^{\sigma}_{j}=0,1$ is the number of fermions with spin $\sigma$ at site $j$, $n_{j}=\sum_{\sigma} n^{\sigma}_{j}$ is the total number of fermions on site $j$, the first $\delta$ is the single-occupancy constraint required by the magnetic moment, and the second $\delta$ means that the total number of fermions (conduction plus Abrikosov fermions) with either spin is $M$.

The target state can be rewritten as a Gutzwiller projected Fermi sea (up to a sign factor) $|\Psi\rangle = P^{\mathrm G}_{0} \prod^{M-1}_{m=0} \prod_{\sigma=\uparrow,\downarrow} \eta^{\dag}_{m,\sigma} |0\rangle$, where $P^{\mathrm G}_{0}$ enforces the single-occupancy constraint and $\eta_{m,\sigma}=\sum^{L}_{j=0} \cos^{m} \theta_{j} \, c_{j,\sigma}$~\footnote{$\cos^{0}\theta_{j}$ is undefined at $\theta_{j}=\pi/2$, but we take it to be $1$ to simplify our notation.} are non-orthogonal unnormalized orbitals constructed from linear combinations of the conduction and Abrikosov fermions~\cite{Supple}. The participation of the Abrikosov fermions in the Fermi sea indicates hybridization between the magnetic moment and the conduction fermions. This second quantized form makes it transparent that $|\Psi\rangle$ is a spin singlet. Gutzwiller projected states have been used extensively in strongly correlated systems~\cite{anderson2004,edegger2007}, and parent Hamiltonians were constructed for some states with an invertible Gutzwiller projector~\cite{laughlin2002,kollar2003}. This is not the case for our model so the previous method cannot be applied directly.

If we introduce another set of operators $\zeta_{0,\sigma} = \sum^{L}_{j=1} c_{j,\sigma}$ and $\zeta_{m,\sigma} = \sum^{L}_{j=1}\cos^{m-1}\theta_{j}(1-\cos\theta_{j})c_{j,\sigma}$ ($m{\geq}1$), the target state can be simplified to~\cite{Supple}
\begin{eqnarray}
|\Psi\rangle &=& \left( c^{\dag}_{0,\uparrow} \zeta^{\dag}_{0,\downarrow} - c^{\dag}_{0,\downarrow} \zeta^{\dag}_{0,\uparrow} \right) \prod^{M-1}_{m=1} \prod_{\sigma=\uparrow,\downarrow} \zeta^{\dag}_{m,\sigma}|0\rangle ,
\label{eq:zetaFS}
\end{eqnarray}
which would be of great help when one searches for the parent Hamiltonian. The Fermi liquid picture of Nozi\`eres is realized in an exact manner: the impurity forms a spin singlet with the $\zeta^{\dag}_{0,\sigma}$ modes and the other conduction fermions form a Fermi sea in the non-orthogonal $\zeta^{\dag}_{m>0,\sigma}$ modes. If the state with a particular $M$ is the ground state, one can add (remove) conduction fermions to (from) the $\zeta^{\dag}_{m>0,\sigma}$ modes to create excited states. However, this procedure can only give us some states in which the numbers of spin-up and spin-down fermions are equal. The Fermi liquid picture in other cases needs to be corroborated using numerical results. For the original Kondo problem, Yosida proposed a trial wave function made of a Kondo singlet and a Fermi sea of conduction fermions~\cite{yosida1966}, but it does not work well~\cite{kondo1967}. The structure of Eq.~(\ref{eq:zetaFS}) is very similar to the proposal of Yosida, but the crucial additional insight is that the single-particle orbitals in the Kondo singlet and the Fermi sea should be chosen properly.

{\em Parent Hamiltonian} --- The target state Eq.~(\ref{eq:wavefunc1}) is the exact ground state of the Hamiltonian $H = \frac{\pi^2}{4L^2} \left( H_{0} + H_{\rm P} + H_{\rm K} + H_{\rm C} \right)$, where
\begin{eqnarray}
H_{0} = \sum^{L-1}_{q=0} \sum_{\sigma=\uparrow,\downarrow} 3q^{2}d^{\dag}_{q,\sigma}d_{q,\sigma}
\label{eq:H0}
\end{eqnarray}
[$d_{q,\sigma}=\sqrt{(1+\delta_{0,q})/L}\sum^{L}_{j=1}\cos(q\theta_{j})c_{j,\sigma}$] describes hopping processes of the conduction fermions,
\begin{eqnarray}
H_{\rm P} = \frac{3}{4} \sum^{L}_{j=1} \sum_{\sigma=\uparrow,\downarrow} \cot^2{\frac{\theta_j}{2}} c^{\dag}_{j,\sigma} c_{j,\sigma}
\label{eq:HP}
\end{eqnarray}
gives site-dependent energies to the conduction fermions, and
\begin{eqnarray}
H_{\rm K} = \sum^{L}_{j=1} \cot^2{\frac{\theta_j}{2}} {\mathbf S}_{0} \cdot {\mathbf S}_{j}
\label{eq:HK}
\end{eqnarray}
stands for long-range spin-exchange interactions between the conduction fermions and the magnetic moment, and $H_{\rm C}=\sum^{L}_{j=1} F(L) c^{\dag}_{j,\sigma} c_{j,\sigma}$ is a $L$-dependent chemical potential [$F(L)=-(3L^2/4-1)$ for odd $L$ and $F(L)=-(3L^2/4-3L/2-1/4)$ for even $L$]. The real space representation of $H_{0}$ contains inverse-square hopping terms and site-dependent potential terms, as manifested by
\begin{eqnarray}
H_{0} &=& \sum^{L}_{j,k=1;j{\neq}k} \left[ \frac{6(-1)^{j-k} }{|z_{j}-z_{k}|^2} - \frac{6(-1)^{j-k} }{|z_{j}-z^{*}_{k}|^2} \right] c^{\dag}_{j,\sigma}c_{k,\sigma} \nonumber \\
&& + \sum^{L}_{j=1} \left[ L^{2}+\frac{1}{2} - \frac{3}{2\sin^{2}\theta_{j}} \right] c^{\dag}_{j,\sigma}c_{j,\sigma},
\end{eqnarray}
where $z_{j}=\exp(i\theta_{j})$ and $z^{*}_{j}=\exp(-i\theta_{j})$ are the complex coordinates of the site $j$ and its mirror image with respect to the real axis [see Fig.~(\ref{Figure1})], respectively~\cite{Supple}. In the limit of $L\rightarrow\infty$ and $j{\ll}L$, the Kondo coupling strength also has an inverse-square form as $\cot^2{\tfrac{\theta_j}{2}}{\sim}(j-1/2)^{-2}$. The ratio between the largest hopping strength and Kondo coupling strength is $3/8$ as $L\rightarrow\infty$. The parent Hamiltonian has three mutually commuting conserved quantities: the total number of fermions $N_{f}=\sum^{L}_{j=0}\sum_{\sigma}c^{\dag}_{j,\sigma}c_{j,\sigma}$, the total spin ${\mathbf S}^2=(\sum^{L}_{j=0}{\mathbf S}_{j})^2$, and its $z$-component $S_{z}=\sum^{L}_{j=0}S^{z}_{j}$.

It is proved in~\cite{Supple} that Eq.~(\ref{eq:zetaFS}) is an exact eigenstate of $H$. The key idea is as follows. When $H_{\rm K}$ is acted on $|\Psi\rangle$, it creates particle-hole excitations (including spin-flipped ones) on top of the Fermi sea. The effect of $H_{\rm P}$ is to compensate ``unwanted'' particle-hole excitations created by $H_{\rm K}$ (similar to the cancellation of diffractive scattering in integrable models solved by Bethe ansatz~\cite{sutherland2004}). The truly remarkable observation is that the remaining particle-hole excitations in $(H_{\rm K}+H_{\rm P})|\Psi\rangle$ can be canceled completely by $H_{0}|\Psi\rangle$, where $H_{0}$ describes conduction fermions with a quadratic dispersion relation and discrete Chebyshev polynomials are their single-particle basis. The eigenvalue of $|\Psi\rangle$ with respect to $H_{0} + H_{\rm P} + H_{\rm K}$ is $(M-1)(2M^{2}-M-3/2)$. This is super-extensive (the leading term scales as $M^{3}$) so the prefactor $\pi^{2}/(4L^{2})$ is adopted to ensure that $H$ is physically valid. It will be demonstrated below that $H_{\rm C}$ selects the half-filled states as the ground states [$M=(L+1)/2$ for odd $L$ and $M=(L+2)/2$ for even $L$].

\begin{figure}
\includegraphics[width=0.48\textwidth]{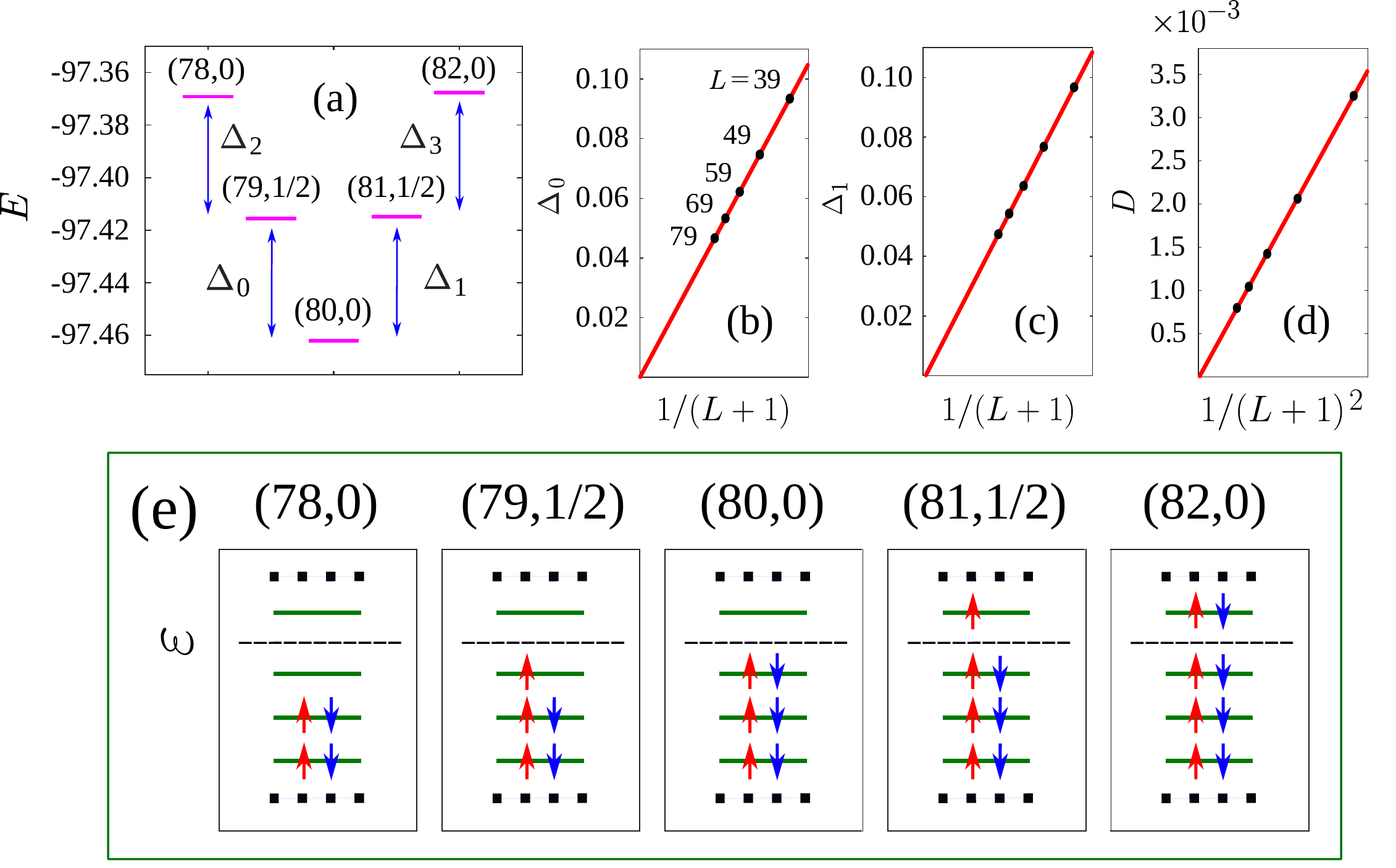}
\caption{(a) Energy spectrum of the $L=79$ system. Each column corresponds to a sector whose quantum numbers are indicated as $(N_{f},S_{z})$ in the panel. (b)-(d) Scaling relations of three quantities in the energy spectrum. (e) Schematics of the Fermi liquid picture for the energy spectrum.}
\label{Figure2}
\end{figure}

{\em Numerical results} --- The proof that $|\Psi\rangle$ is an eigenstate of $H$ is not sufficient because we have claimed that it is actually the ground state. To this end, exact diagonalizations are performed for various choices of $L,N_{f},S_{z}$ with $L$ up to $14$. For all the cases with even $N_{f}$ and $S_{z}=0$, the overlaps between numerically generated lowest eigenstates and Eq.~(\ref{eq:wavefunc1}) are exactly $1$ (up to machine precision). The half-filled states also turn out to be the ground states. The density matrix renormalization group (DMRG)~\cite{White1992,Schollwock2011,Hubig2015,Hubig2017} method is employed to compute the lowest eigenstates for various choices of $L,N_{f},S_{z}$ with $L$ up to $80$. The lowest energy of the $(L,N_{f},S_{z})$ sector is denoted as $E(L,N_{f},S_{z})$. The accuracy of our approach can be checked by comparing numerical and analytical values of $E(L,N_{f},0)$ with even $N_{f}$. For instance, our DMRG results at $L=79$ and $N_{f}=78,80,82$ have absolute errors of the $10^{-7}$ order, so they are almost exact. DMRG calculations also confirm that the half-filled states are the ground states.

The low-lying energy levels of the $L=79$ system are presented and analyzed in Fig.~\ref{Figure2}. The differences in energy eigenvalues are characterized by $\Delta_{0}=E(L,L,1/2)-E(L,L+1,0)$, $\Delta_{1}(L)=E(L,L+2,1/2)-E(L,L+1,0)$, and $D(L)=\Delta_{1}(L)-\Delta_{0}(L)=E(L,L+2,1/2)-E(L,L,1/2)$. The finite-size scaling analysis in Fig.~\ref{Figure2} (b)-(c) reveals that (i) $\Delta_{0}$ and $\Delta_{1}$ go to zero as $1/(L+1)$, and (ii) $D(L)$ go to zero as $1/(L+1)^2$. The same $L$ dependence is also observed in $\Delta_{2}=E(L,L-1,0)-E(L,L+1,0),\Delta_{3}(L)=E(L,L+3,0)-E(L,L+1,0)$, and $\Delta_{3}(L)-\Delta_{2}(L)$. The energy spectrum can be interpreted using the Fermi liquid picture of Nozi\`eres as illustrated in Fig.~\ref{Figure2} (e): there is a Fermi level at $\varepsilon_{0}=0$, the single-particle energy levels are equally spaced at $\varepsilon_{m}=\pm\tfrac{2{\pi}v_{F}}{L+1}(m-\tfrac12)$ ($m=1,2,\cdots$ and $v_{F}$ is the Fermi velocity), the ground state is constructed by filling all the single-particle states below the Fermi level, and the excited states are obtained by adding and/or removing some fermions from the ground state.

\begin{figure}
\includegraphics[width=0.48\textwidth]{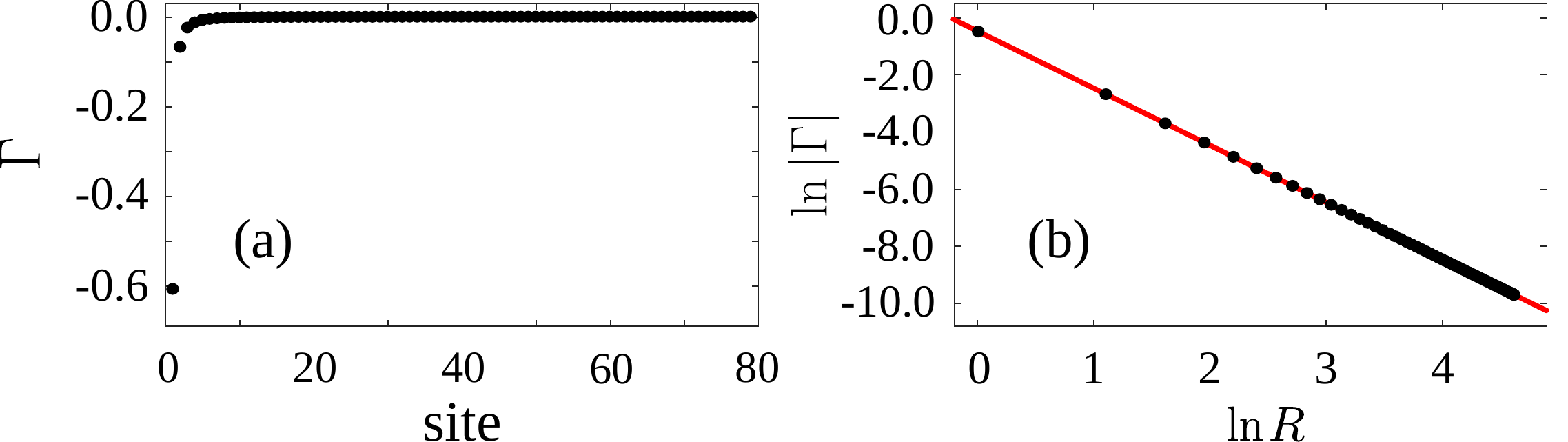}
\caption{Spin-spin correlation function $\Gamma$ of the $L=79$ system. (a) $\Gamma_{j}$ on all sites. (b) Log-log plot of $|\Gamma_{j}|$ versus $R_{j}=\tfrac{2(L+1)}{\pi}\sin\tfrac{\theta_{j}}{2}$.}
\label{Figure3}
\end{figure}

The spin-spin correlation function $\Gamma_{j}=\langle\Psi| {\mathbf S}_{0} \cdot {\mathbf S}_{j} |\Psi\rangle/\langle\Psi|\Psi\rangle$ for the $L=79$ system is shown in Fig.~\ref{Figure3}~\cite{Supple}. As an important signature of the Kondo physics, the decaying behavior of $\Gamma_{j}$ has been studied extensively using CFT and numerical methods~\cite{barzykin1998,hand2006,borda2007,holzner2009}. One generally observes a crossover from $1/r$ decay in the Kondo screening cloud to $1/r^{2}$ decay outside the cloud. The latter is easy to understand because the system behaves essentially as free fermions once the impurity is completely screened. For our model, the distance between the sites $0$ and $j$ is defined as $R_{j}=\tfrac{2(L+1)}{\pi}\sin\tfrac{\theta_{j}}{2}$~\footnote{This is the case if we change the radius to $\tfrac{L+1}{\pi}$ in Fig.~\ref{Figure1} (a) and use the chord distance between two sites. It ensures that $R_{j}{\approx}j$ when $j{\ll}L$. The linear coordinates $u_{j}$ are rescaled but the wave function is only modified in a trivial way.}. The log-log plot of $|F_{j}|$ versus $R_{j}$ in Fig.~\ref{Figure3} (b) has a slope $-2.000$ with a coefficient of determination higher than $99.99\%$. This means that the Kondo screening length is extremely short so we can only see the free fermion decay.

\begin{figure}
\includegraphics[width=0.48\textwidth]{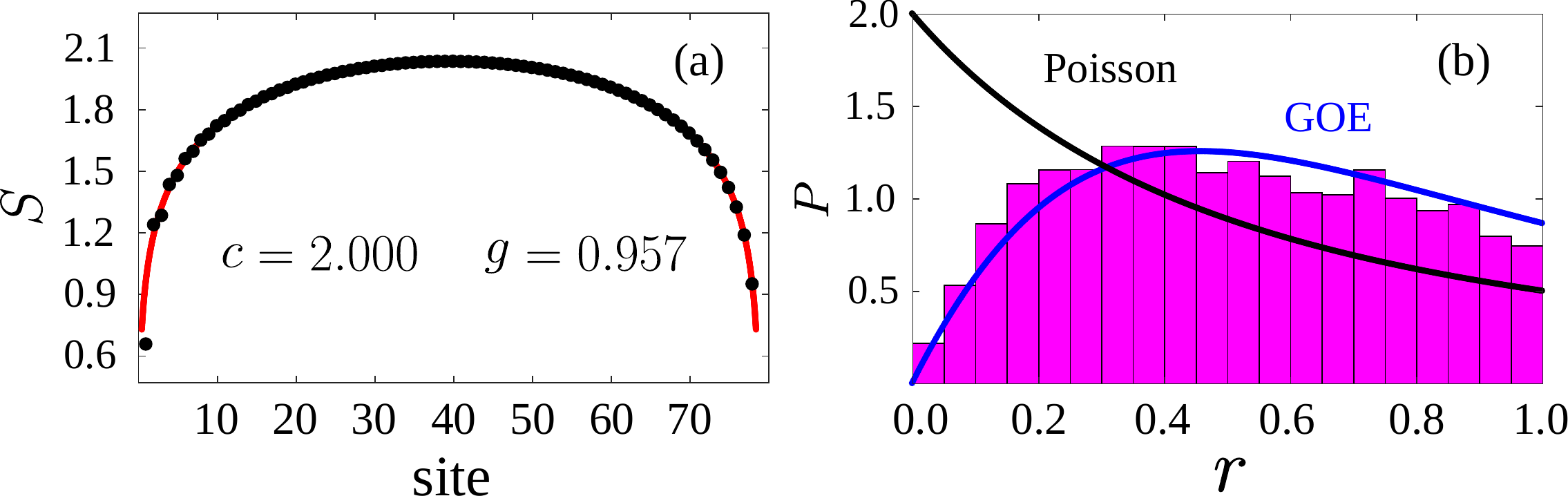}
\caption{(a) von Neumann entanglement entropy of the ground state of the $L=79, N_{f}=80, S_{z}=0$ system. The red line is the least square fit using the data points in $10{\leq}L_{A}{\leq}70$. (b) Energy level statistics of the $L=9$ system in the subspace with $N_{f}=10,S_{z}=0,{\mathbf S}^2=0$. The bars are the probability density of the ratio $r_{m}$ and the lines are the Poisson and GOE results.}
\label{Figure4}
\end{figure}

The von Neumann entanglement entropy has also been studied. It is defined as $S(L_{A})=-{\rm Tr}(\rho_{A}\ln\rho_{A})$ with $\rho_{A}$ being the reduced density matrix for the subsystem $0{\leq}j{\leq}L_{A}$. It is expected that the Calabrese-Cardy formula
\begin{eqnarray}
S(L_{A}) = \frac{c}{6} \ln \left[ \frac{L}{\pi} \sin\left( \pi \frac{L_{A}}{L} \right) \right] + g
\end{eqnarray}
is still applicable outside the Kondo screening cloud~\cite{calabrese2004,affleck2009}, where $c$ is the CFT central charge and $g$ is a non-universal number. For the $L=79,N_{f}=80,S_{z}=0$ case, the data points in the range $10{\leq}L_{A}{\leq}70$ can be fitted using $c=2.000$ and $g=0.957$ as shown in Fig.~\ref{Figure4} (a). This is consistent with the fact that the system contains two gapless modes. For the first few sites, there are obvious deviations from the Calabrese-Cardy formula, which should reflect certain aspects of the coupling between the impurity and conduction fermions, but it is unclear how to extract such information.

It is natural to ask if our model is integrable given that a large class of inverse-square models, such as the celebrated Haldane-Shastry model, has been proved to be integrable~\cite{haldane1988,shastry1988,haldane1992,bernard1993}. One diagnostic of integrability is level statistics of the energy spectrum~\cite{poilblanc1993,Atas2013}. The eigenvalues $E_{m}$ are sorted according to the conserved quantum numbers and arranged in ascending order, then the level spacing $\delta_{m}=E_{m+1}-E_{m}$ and the ratio $r_{m}=\min(\delta_{m},\delta_{m+1})/\max(\delta_{m},\delta_{m+1})$ can be obtained. The probability density of $r_{m}$ have two possible forms: a Poisson distribution with $P(r)=2/(1+r)^2$ for integrable models and a Gaussian orthogonal ensemble (GOE) with $P(r)=[27(r+r^2)]/[4(1+r+r^2)^{5/2}]$ for non-integrable models. It is quite plausible that our model obeys the GOE distribution as shown in Fig.~\ref{Figure4} (b). The expectation value of $r_{m}$ is $0.511$, quite close to the theoretical value $0.536$ for GOE. This suggests that our model is not integrable despite having explicit forms for the ground states in all sectors with even $N_f$.

{\em Conformal field theory formulation} --- The successful construction of an exactly solvable model for the spin-1/2 Kondo problem with polynomial wave function motivates us to search for similar scenarios in more complicated quantum impurity systems, such as the multichannel and SU(N) Kondo problems. A promising route along this direction manifests itself as soon as we establish a link between the wave function and the associated boundary CFT, which is very similar to the practice of constructing quantum Hall wave functions from edge CFT~\cite{moore1991}. The key observation is that the wave function (\ref{eq:wavefunc2}) can be represented as a CFT correlator~\cite{Supple}
\begin{eqnarray}
&& \Psi(\{n^\uparrow_{j}\},\{n^\downarrow_{j}\}) = \nonumber \\
&& \quad \langle \mathcal{O}_{\mathrm{bg}} A^{n^\uparrow_{0},n^\downarrow_{0}}(u_{0}) A^{n^\uparrow_{1},n^\downarrow_{1}}(u_{1}) \cdots A^{n^\uparrow_{L},n^\downarrow_{L}}(u_{L}) \rangle
\label{eq:iMPS}
\end{eqnarray}
with a background charge $\mathcal{O}_{\mathrm{bg}}$ controlling the total number of fermions and vertex operators
\begin{equation}
A^{n^\uparrow_{j},n^\downarrow_{j}}(u_{j})=\left\{
\begin{array}{c}
\delta_{n_{j},1} :e^{i\sum_{\sigma} n^\sigma_{j} \phi_{\sigma}(u_{j})}:  \quad j=0 \\
:e^{i\sum_{\sigma} n^\sigma_{j}  \phi_{\sigma}(u_{j})}:  \quad j=1,\cdots,L
\end{array}
\right. ,
\label{eq:vertexops}
\end{equation}
where $\phi_{\sigma}(u_{j})$ is a two-component chiral bosonic field and $:\cdots:$ denotes normal ordering.

This reformulation can be viewed as an infinite-dimensional matrix product state~\cite{cirac2010} and turns out to be very illuminating in further analysis. If the vertex operators for the impurity site were removed from Eq.~(\ref{eq:iMPS}), the wave function reduces to the ground state of $H_{0}$ that realizes the free-fermion CFT with the free boundary condition on the lattice~\cite{tu2015,basumallick2016,stephan2017,hackenbroich2017}. This is perhaps not too surprising, since the vertex operators (\ref{eq:vertexops}) for conduction fermions all have conformal weight $h=1/2$ and thus are fermionic fields. However, from the boundary CFT perspective, it is known that the impurity site plays the role of boundary condition changing (BCC) operator as it changes the free boundary condition of conduction fermions to the ``Kondo boundary condition"~\cite{affleck1994}. One may speculate that the vertex operator on the impurity site in Eq.~(\ref{eq:iMPS}) acts as a BCC operator which changes the free boundary condition at one end of the chain to the Kondo boundary condition. Furthermore, the fact that the vertex operator at the impurity site [first line of Eq.~(\ref{eq:vertexops})] is also a fermionic field with conformal weight $h=1/2$ suggests that the Fermi liquid picture holds. The identification of our wave function as a CFT correlator would shed light on how to design other models with CFT inspired wave functions.

{\em Conclusion and discussion} --- In summary, we have constructed an exactly solvable quantum impurity model with a highly structured ground state and a parent Hamiltonian consists of inverse-square hopping and interaction terms. The low-energy physics of our model is described by a boundary CFT and a particular correlator of this CFT reproduces the wave function. This work establishes a connection between the wave function and boundary CFT in quantum impurity models, which paves the way toward constructing wave functions and exactly solvable models for more complicated systems with SU(N) fermions and/or multiple impurities~\cite{wu2019}. For Kondo problems that cannot be solved exactly, we may define $\zeta^{\dag}$ modes with unknown parameters and use Eq.~(\ref{eq:zetaFS}) as variational ansatz. The numerical prospect of optimizing these parameters with respect to some given Hamiltonians is left for future works.

{\em Acknowledgment} --- We are grateful to Jan von Delft, Biao Huang, Seung-Sup Lee, Fr\'ed\'eric Mila, Germ\'an Sierra, Andreas Weichselbaum, and Guang-Ming Zhang for helpful discussions. This work was supported by the DFG via project A06 (HHT) of SFB 1143 (project-id 247310070), the NSFC under Grant No. 11804107 (YHW), and the startup grant of HUST (YHW).

\bibliography{Kondo}

\clearpage

\setcounter{figure}{0}
\setcounter{equation}{0}
\renewcommand\thefigure{A\arabic{figure}}
\renewcommand\theequation{A\arabic{equation}}

\begin{widetext}

\section{Appendix A: Wave Function}

This section provides more details about the properties of the ground-state wave function
\begin{eqnarray}
\Psi(\{n^\uparrow_{j}\},\{n^\downarrow_{j}\}) &=& \delta_{n_{0}=1} \delta_{\sum_{j} n^{\uparrow}_{j} = \sum_{j} n^{\downarrow}_{j} = M} \prod_{\sigma=\uparrow,\downarrow} \prod_{0{\leq}j<k{\leq}L} (u_{j}-u_{k})^{n^\sigma_{j}n^\sigma_{k}} .
\label{eq:app-wavefunc}
\end{eqnarray}

\subsection{A1: Gutzwiller projected Fermi sea}

We first prove Eq.~(3) of the main text. To begin with, we can rewrite the ground state as
\begin{eqnarray}
|\Psi\rangle &=& \sum_{\{n^\uparrow_{j}\},\{n^\downarrow_{j}\}} \Psi(\{n^\uparrow_{j}\},\{n^\downarrow_{j}\}) \,
             (c^{\dag}_{0,\uparrow})^{n^\uparrow_0} \cdots (c^{\dag}_{L,\uparrow})^{n^\uparrow_{L}}
             (c^{\dag}_{0,\downarrow})^{n^\downarrow_{0}} \cdots (c^{\dag}_{L,\downarrow})^{n^\downarrow_{L}} |0\rangle = P^{\mathrm G}_{0} |{\widetilde\Psi}\rangle.
\label{eq:app-gutz1}
\end{eqnarray}
The Gutzwiller projector $P^{\mathrm G}_j$ enforces the single-occupancy constraint $\delta_{{n_0}=1}$ on the impurity site $j=0$ and the state $|{\widetilde\Psi}\rangle$ is
\begin{eqnarray}
&& \sum_{\{n^\uparrow_{j}\},\{n^\downarrow_{j}\}} \delta_{\sum_{j} n^{\uparrow}_{j} = \sum_j n^{\downarrow}_{j} = M}
   \left[ \prod_{\sigma=\uparrow,\downarrow} \prod_{0{\leq}j<k{\leq}L} (u_{j}-u_{k})^{n^\sigma_{j} n^\sigma_{k}} \right] \, (c^{\dag}_{0,\uparrow})^{n^\uparrow_{0}} \cdots (c^{\dag}_{L,\uparrow})^{n^\uparrow_{L}} (c^{\dag}_{0,\downarrow})^{n^\downarrow_{0}} \cdots (c^{\dag}_{L,\downarrow})^{n^\downarrow_{L}} |0\rangle \nonumber \\
          = && \sum_{x_1<\cdots<x_{M}} \sum_{y_1<\cdots<y_{M}} \left[ \prod_{1{\leq}j<k{\leq}M} (u_{x_{j}}-u_{x_{k}}) \prod_{1{\leq}j<k{\leq}M} (u_{y_{j}}-u_{y_{k}}) \right] \, c^{\dag}_{x_{1},\uparrow} \cdots c^{\dag}_{x_{M},\uparrow} c^{\dag}_{y_{1},\downarrow} \cdots c^{\dag}_{y_{M},\downarrow} |0\rangle,
\label{eq:app-gutz2}
\end{eqnarray}
where $\{x_{1},\cdots,x_{M}\}$ and $\{y_{1},\cdots,y_{M}\}$ are the lattice sites occupied by spin-up and spin-down fermions, respectively. The Jastrow factor $\prod_{1{\leq}j<k{\leq}M} (u_{x_{j}}-u_{x_{k}})$ can be converted to a Vandermonde determinant
\begin{equation}
(-1)^{\frac{1}{2}M(M-1)}\det
\begin{pmatrix}
1 & 1 & \cdots & 1 \\
u_{x_{1}} & u_{x_{2}} & \cdots & u_{x_{M}} \\
\vdots & \vdots & \ddots & \vdots \\
u^{M-1}_{x_{1}} & u^{M-1}_{x_{2}} & \cdots & u^{M-1}_{x_{M}} \\
\end{pmatrix},
\end{equation}
so one can see that
\begin{eqnarray}
|{\widetilde\Psi}\rangle &\propto& \sum_{\{x_i\},\{y_i\}} \det
\begin{pmatrix}
1 & 1 & \cdots & 1 \\
u_{x_{1}} & u_{x_{2}} & \cdots & u_{x_{M}} \\
\vdots & \vdots & \ddots & \vdots \\
u^{M-1}_{x_{1}} & u^{M-1}_{x_{2}} & \cdots & u^{M-1}_{x_{M}} \\
\end{pmatrix}
\det
\begin{pmatrix}
1 & 1 & \cdots & 1 \\
u_{y_{1}} & u_{y_{2}} & \cdots & u_{y_{M}} \\
\vdots & \vdots & \ddots & \vdots \\
u^{M-1}_{y_{1}} & u^{M-1}_{y_{2}} & \cdots & u^{M-1}_{y_{M}} \\
\end{pmatrix}
\, c^{\dag}_{x_{1},\uparrow} \cdots c^{\dag}_{x_{M},\uparrow} c^{\dag}_{y_{1},\downarrow} \cdots c^{\dag}_{y_{M},\downarrow} |0\rangle \nonumber \\
&\propto& \prod^{M-1}_{m=0} \prod_{\sigma=\uparrow,\downarrow} \eta^{\dag}_{m,\sigma} |0\rangle,
\label{eq:app-gutz3}
\end{eqnarray}
with $\eta_{m,\sigma}=\sum^{L}_{j=0} u^{m}_{j} \, c_{j,\sigma}$. The substitution of Eq.~(\ref{eq:app-gutz3}) into Eq.~(\ref{eq:app-gutz1}) completes the proof.

The Gutzwiller projection can be performed explicitly to facilitate our construction of the parent Hamiltonian. For this purpose, we use the linear combinations of the $\eta$ modes to define some $\zeta$ modes
\begin{eqnarray}
\zeta_{0,\sigma} = \eta_{0,\sigma}-c_{0,\sigma} = \sum^{L}_{j=1} c_{j}, \;\;\;\;\;\; \zeta_{m,\sigma} = \eta_{m-1,\sigma}-\eta_{m,\sigma} = \sum^{L}_{j=1}\cos^{m-1}\theta_{j}(1-\cos\theta_{j})c_{j,\sigma} \;\;\; \text{for} \;\;\; m{\geq}1 \; .
\end{eqnarray}
For odd $L$, it is possible to have $\theta_{j}=\pi/2$ such that $\cos^{m-1}\theta_{j}$ is undefined at $m=1$. For these cases, we define $\cos^{0}\theta_{j}{\equiv}1$ to simplify our notation. The ground state is transformed to
\begin{eqnarray}
|\Psi\rangle &=& P^{\rm G}_{0} \eta^{\dag}_{0,\uparrow} \eta^{\dag}_{0,\downarrow} \prod^{M-1}_{m=1} \prod_{\sigma=\uparrow,\downarrow} \eta^{\dag}_{m,\sigma} |0\rangle \notag \\
 &=& P^{\rm G}_{0} \left( c^{\dag}_{0,\uparrow}+\zeta^{\dag}_{0,\uparrow} \right) \left( c^{\dag}_{0,\downarrow}+\zeta^{\dag}_{0,\downarrow} \right) \prod^{M-1}_{m=1} \prod_{\sigma=\uparrow,\downarrow} \zeta^{\dag}_{m,\sigma} |0\rangle \notag \\
&=& \left( c^{\dag}_{0,\uparrow}\zeta^{\dag}_{0,\downarrow}-c^{\dag}_{0,\downarrow}\zeta^{\dag}_{0,\uparrow} \right) \prod^{M-1}_{m=1} \prod_{\sigma=\uparrow,\downarrow} \zeta^{\dag}_{m,\sigma}|0\rangle.
\end{eqnarray}

\subsection{A3: Conformal field theory formulation}

This subsection aims to prove Eq.~(10) of the main text. The chiral bosonic field for spin-up fermions is defined by $\phi_{\uparrow}(u)=\phi^{\uparrow}_{0}-i\pi^{\uparrow}_{0}\ln u+i\sum^{\infty}_{n\neq 0}\frac{1}{n}a^{\uparrow}_{n}u^{-n}$, where $\phi^{\uparrow}_{0}$, $\pi^{\uparrow}_{0}$, and $a^{\uparrow}_{n}$ are operators satisfying the following commutation relations: $[\phi^{\uparrow}_{0}, \pi^{\uparrow}_{0}]=i$ and $[a^{\uparrow}_{n},a^{\uparrow}_{m}]=n\delta_{n+m,0}$. Furthermore, $\pi^{\uparrow}_{0}$ and $a^{\uparrow}_{n>0}$ annihilate the vacuum $|0\rangle$, i.e., $\pi^{\uparrow}_{0} |0\rangle = a^{\uparrow}_{n>0}|0\rangle = 0$. The corresponding chiral vertex operator for spin-up fermion is given by
\begin{eqnarray}
:\exp\left[i\phi_{\uparrow}(u)\right]: = \exp\left(i\phi^{\uparrow}_{0}+\sum^{\infty}_{n=1}\frac{1}{n}a^{\uparrow}_{-n}u^{n}\right)
\exp\left(\pi^{\uparrow}_{0}\ln u-\sum^{\infty}_{n=1}\frac{1}{n}a^{\uparrow}_{n}u^{-n}\right).
\label{eq:app-wavefunc1}
\end{eqnarray}
For odd $L$, $\theta_{j=(L+1)/2} = \pi/2$ and, for this site, the corresponding vertex operator only includes the zero-mode part $\exp(i\phi^{\uparrow}_{0})$. For spin-down fermions, the expressions of the chiral bosonic field and vertex operators are similar, with the spin index replaced by $\downarrow$.

The product of chiral vertex operators is obtained by normal ordering of operators
\begin{eqnarray}
&\phantom{=}& : \exp\left[ i\phi_{\uparrow}(u_{x_{1}}) \right] : \cdots : \exp\left[ i\phi_{\uparrow}(u_{x_{M}}) \right] :  \nonumber \\
&=& \exp\left( iM\phi^{\uparrow}_{0} + \sum^{M}_{m=1} \sum^{\infty}_{n=1} \frac{1}{n} a^{\uparrow}_{-n} u^{n}_{x_{m}}\right) \exp\left[ \pi^{\uparrow}_{0}\ln(u_{x_{1}}{\cdots}u_{x_{M}}) - \sum^{M}_{m=1} \sum^{\infty}_{n=1} \frac{1}{n} a^{\uparrow}_{n} u^{-n}_{x_{m}} \right] \nonumber \\
&\phantom{=}& \times \prod_{1{\leq}i<j{\leq}M} \left( u_{x_{i}}-u_{x_{j}} \right).
\label{eq:app-wavefunc2}
\end{eqnarray}
The electron vertex operators can be neutralized using a background charge operator
\begin{eqnarray}
O^{\uparrow}_{bg} = \exp\left(-iM\phi^{\uparrow}_{0}\right),
\end{eqnarray}
so that the following chiral correlator (expectation value in the CFT vacuum $|0\rangle$) yields the Jastrow product for spin-up fermions
\begin{eqnarray}
\langle O^{\uparrow}_{bg} :\exp\left[i\phi_{\uparrow}(u_{x_{1}})\right]: \cdots :\exp\left[i\phi_{\uparrow}(u_{x_{M}})\right]: \rangle = \prod_{1{\leq}i<j{\leq}M} \left( u_{x_{i}}-u_{x_{j}} \right).
\end{eqnarray}
For the wave function with both spin-up and spin-down fermions being present, the background charge should be defined as $O_{bg}=O^{\uparrow}_{bg}O^{\downarrow}_{bg}$.

\section{Appendix B: Parent Hamiltonian}

This section provides details about how to act the parent Hamiltonian on the ground state. The target state is a superposition of two parts in which the spin on the $j=0$ site points up and down, respectively. When acted upon by an operator, the result can still be written like this. To simplify subsequent analysis, we focus on the spin-up part. The spin-down part can be analyzed in the same way. For notational ease, we denote $\prod^{M-1}_{m=1} \prod_{\sigma=\uparrow,\downarrow} \zeta^{\dag}_{m,\sigma}|0\rangle$ as $|{\rm PFS}\rangle$ (PFS stands for partial Fermi sea).

\subsection{B1: The Kondo term}

The Kondo term is
\begin{eqnarray}
H_{\rm K} = \mathbf{S}_{0} \cdot \Lambda_{\rm K} = \frac{1}{2} S^{+}_{0} \Lambda^{-}_{\rm K} + \frac{1}{2} S^{-}_{0} \Lambda^{+}_{\rm K} + S^{z}_{0} \Lambda^{z}_{\rm K}
\end{eqnarray}
with
\begin{eqnarray}
\Lambda_{\rm K} &=& \sum^{L}_{j=1} \frac{1+\cos\theta_{j}}{1-\cos\theta_{j}} \mathbf{S}_{j} = \sum_{j=1}^{L}\cot^{2}\frac{\theta_{j}}{2} \mathbf{S}_{j}.
\end{eqnarray}
As for usual spin operators, one can also define $\Lambda^{+}_{\rm K}$, $\Lambda^{-}_{\rm K}$, and $\Lambda^{z}_{\rm K}$. Their commutators with the $\zeta$ modes are
\begin{eqnarray}
&& \left[ \Lambda^{-}_{\rm K} , \zeta^{\dag}_{0,\uparrow} \right] = \sum^{L}_{j=1} \frac{1+\cos\theta_{j}}{1-\cos\theta_{j}} c^{\dag}_{j,\downarrow} \equiv O_{{\rm K},\downarrow}, \;\;\; \left[ \Lambda^{-}_{\rm K} , \zeta^{\dag}_{m,\uparrow} \right] = 2\zeta^{\dag}_{0,\downarrow} - 2 \sum^{m-1}_{\alpha=1} \zeta^{\dag}_{\alpha,\downarrow} - \zeta^{\dag}_{m,\downarrow} \;\;\; \text{for} \;\;\; m>0, \notag \\
&& \left[ \Lambda^{-}_{\rm K} , \zeta^{\dag}_{m,\downarrow} \right] = 0 \;\;\; \text{for} \;\;\; m{\geq}0, \notag \\
&& \left[ \Lambda^{z}_{\rm K} , \zeta^{\dag}_{0,\uparrow} \right] = \frac{1}{2} \sum^{L}_{j=1} \frac{1+\cos\theta_{j}}{1-\cos\theta_{j}} c^{\dag}_{j,\uparrow} \equiv \frac{1}{2} O_{{\rm K},\uparrow}, \;\;\; \left[ \Lambda^{z}_{\rm K} , \zeta^{\dag}_{m,\uparrow} \right] = \zeta^{\dag}_{0,\uparrow} - \sum^{m-1}_{\alpha=1} \zeta^{\dag}_{\alpha,\uparrow} - \frac{1}{2} \zeta^{\dag}_{m,\uparrow} \;\;\; \text{for} \;\;\; m>0, \notag \\
&& \left[ \Lambda^{z}_{\rm K} , \zeta^{\dag}_{0,\downarrow} \right] = -\frac{1}{2} \sum^{L}_{j=1} \frac{1+\cos\theta_{j}}{1-\cos\theta_{j}} c^{\dag}_{j,\downarrow} = -\frac{1}{2} O_{{\rm K},\downarrow}, \;\;\; \left[ \Lambda^{z}_{\rm K} , \zeta^{\dag}_{m,\downarrow} \right] = -\zeta^{\dag}_{0,\downarrow} + \sum^{m-1}_{\alpha=1} \zeta^{\dag}_{\alpha,\downarrow} + \frac{1}{2} \zeta^{\dag}_{m,\downarrow} \;\;\; \text{for} \;\;\; m>0.
\end{eqnarray}
The action of $H_{\rm K}$ on $|\Psi\rangle$ gives $|\Psi^{\uparrow}_{\rm K}\rangle+|\Psi^{\downarrow}_{\rm K}\rangle$ with
\begin{eqnarray}
|\Psi^{\uparrow}_{\rm K}\rangle = c^{\dag}_{0,\uparrow} \left( -\frac{1}{2} \Lambda^{-}_{\rm K} \zeta^{\dag}_{0,\uparrow} + \frac{1}{2} \Lambda^{z}_{\rm K} \zeta^{\dag}_{0,\downarrow} \right) |{\rm PFS}\rangle = c^{\dag}_{0,\uparrow} \left( -\frac{3}{4} O_{{\rm K},\downarrow} - \frac{1}{2} \zeta^{\dag}_{0,\uparrow} \Lambda^{-}_{\rm K} + \frac{1}{2} \zeta^{\dag}_{0,\downarrow} \Lambda^{z}_{\rm K} \right) |{\rm PFS}\rangle.
\label{eq:kondo-spin-up}
\end{eqnarray}
The commutators listed above can be used to show that
\begin{eqnarray}
\zeta^{\dag}_{0,\uparrow} \Lambda^{-}_{\rm K} |{\rm PFS}\rangle &=& 2\sum^{M-1}_{m=1} \zeta^{\dag}_{0,\uparrow} \zeta^{\dag}_{0,\downarrow} \left[ \prod^{m-1}_{\alpha=1} \zeta^{\dag}_{\alpha,\uparrow} \zeta^{\dag}_{\alpha,\downarrow} \right] \zeta^{\dag}_{m,\downarrow} \left[ \prod^{M-1}_{\alpha=m+1} \zeta^{\dag}_{\alpha,\uparrow} \zeta^{\dag}_{\alpha,\downarrow} \right] |0\rangle, \notag \\
\zeta^{\dag}_{0,\downarrow} \Lambda^{z}_{\rm K} |{\rm PFS}\rangle &=& -\sum^{M-1}_{m=1} \zeta^{\dag}_{0,\uparrow} \zeta^{\dag}_{0,\downarrow} \left[ \prod^{m-1}_{\alpha=1} \zeta^{\dag}_{\alpha,\uparrow} \zeta^{\dag}_{\alpha,\downarrow} \right] \zeta^{\dag}_{m,\downarrow} \left[ \prod^{M-1}_{\alpha=m+1} \zeta^{\dag}_{\alpha,\uparrow} \zeta^{\dag}_{\alpha,\downarrow} \right] |0\rangle.
\label{eq:kondo-unwanted}
\end{eqnarray}
The terms in Eq.~(\ref{eq:kondo-spin-up}) are unwanted in the sense that they do not appear in the target state. However, the result is not bad because the unwanted terms in Eq.~(\ref{eq:kondo-unwanted}) are highly structured. The $O_{{\rm K},\downarrow}$ term is somewhat special, so our next step would be to eliminate it.

\subsection{B2: The site-dependent potential term}

The site-dependent potential term is
\begin{eqnarray}
H_{\rm P} = \frac{3}{4} \sum^{L}_{j=1} \sum_{\sigma=\uparrow,\downarrow} \frac{1+\cos\theta_{j}}{1-\cos\theta_{j}} c^{\dag}_{j,\sigma} c_{j,\sigma} = \frac{3}{4} \sum^{L}_{j=1} \sum_{\sigma=\uparrow,\downarrow} \cot^{2}\frac{\theta_{j}}{2} c^{\dag}_{j,\sigma} c_{j,\sigma}.
\end{eqnarray}
Its commutators with the $\zeta$ modes are
\begin{eqnarray}
 \left[ H_{\rm P} , \zeta^{\dag}_{0,\sigma} \right] = \frac{3}{4} \sum^{L}_{j=1} \frac{1+\cos\theta_{j}}{1-\cos\theta_{j}} c^{\dag}_{j,\sigma} = \frac{3}{4} O_{{\rm K},\sigma}, \;\;\; \left[ H_{\rm P} , \zeta^{\dag}_{m,\sigma} \right] = \frac{3}{2} \zeta^{\dag}_{0,\sigma} - \frac{3}{2} \sum^{m-1}_{\alpha=1} \zeta^{\dag}_{\alpha,\sigma} - \frac{3}{4} \zeta^{\dag}_{m,\sigma} \;\;\; \text{for} \;\;\; m>0.
\end{eqnarray}
The action of $H_{\rm P}$ on $|\Psi\rangle$ gives $|\Psi^{\uparrow}_{\rm P}\rangle+|\Psi^{\downarrow}_{\rm P}\rangle$ with
\begin{eqnarray}
|\Psi^{\uparrow}_{\rm P}\rangle = c^{\dag}_{0,\uparrow} H_{\rm P} \zeta^{\dag}_{0,\downarrow} |{\rm PFS}\rangle = c^{\dag}_{0,\uparrow} \left( \frac{3}{4} O_{{\rm K},\downarrow} + \zeta^{\dag}_{0,\downarrow} H_{\rm P} \right) |{\rm PFS}\rangle .
\label{eq:chemical-spin-up}
\end{eqnarray}
The commutators listed above can be used to show that
\begin{eqnarray}
\zeta^{\dag}_{0,\downarrow} H_{\rm P} |{\rm PFS}\rangle = -\frac{3}{2} \sum^{M-1}_{m=1} \zeta^{\dag}_{0,\uparrow} \zeta^{\dag}_{0,\downarrow} \left[ \prod^{m-1}_{\alpha=1} \zeta^{\dag}_{\alpha,\uparrow} \zeta^{\dag}_{\alpha,\downarrow} \right] \zeta^{\dag}_{m,\downarrow} \left[ \prod^{M-1}_{\alpha=m+1} \zeta^{\dag}_{\alpha,\uparrow} \zeta^{\dag}_{\alpha,\downarrow} \right] |0\rangle - \frac{3}{2} (M-1) \zeta^{\dag}_{0,\downarrow} |{\rm PFS}\rangle.
\label{eq:chemical-unwanted}
\end{eqnarray}
At this point, one can see why this choice of $H_{\rm P}$ is promising: the $O_{{\rm K},\downarrow}$ terms in Eqs.~(\ref{eq:kondo-spin-up}) and~(\ref{eq:chemical-spin-up}) cancel each other; the unwanted terms in Eq.~(\ref{eq:chemical-unwanted}) have the same form as those in Eq.~(\ref{eq:kondo-unwanted}). It is obvious that $|\Psi^{\uparrow}_{\rm K}\rangle+|\Psi^{\uparrow}_{\rm P}\rangle$ is
\begin{eqnarray}
c^{\dag}_{0,\uparrow} \left\{ -3 \sum^{M-1}_{m=1} \zeta^{\dag}_{0,\uparrow} \zeta^{\dag}_{0,\downarrow} \left[ \prod^{m-1}_{\alpha=1} \zeta^{\dag}_{\alpha,\uparrow} \zeta^{\dag}_{\alpha,\downarrow} \right] \zeta^{\dag}_{m,\downarrow} \left[ \prod^{M-1}_{\alpha=m+1} \zeta^{\dag}_{\alpha,\uparrow} \zeta^{\dag}_{\alpha,\downarrow} \right] |0\rangle \right\} -\frac{3}{2} (M-1) c^{\dag}_{0,\uparrow} \zeta^{\dag}_{0,\downarrow} |{\rm PFS}\rangle ,
\label{eq:kondo-chemical-unwanted}
\end{eqnarray}
where the first term represents particle-hole excitations on top of the Fermi sea. In order to find a parent Hamiltonian, these particle-hole excitations should be canceled by contributions from $H_{\rm 0} |{\rm PFS}\rangle$.

\subsection{B3: The hopping term}

The hopping term is designed as
\begin{eqnarray}
H_{0} = \sum^{L-1}_{q=0} \sum_{\sigma=\uparrow,\downarrow} 3q^{2} d^{\dag}_{q,\sigma} d_{q,\sigma},
\end{eqnarray}
where
\begin{eqnarray}
d_{q,\sigma}=
\begin{cases}
\frac{1}{\sqrt{L}} \sum^{L}_{j=1} c_{j,\sigma} \quad \text{for} \quad q=0 \\
\sqrt{\frac{2}{L}} \sum^{L}_{j=1} \cos(q\theta_{j})c_{j,\sigma} \quad \text{for} \quad 1{\leq}q{\leq}L-1
\end{cases}.
\end{eqnarray}
The angles are chosen to be $\theta_{j}=\frac{\pi}{L}(j-\frac{1}{2})$ to make sure that the fermionic modes $d_{q,\sigma}$ form an {\rm orthonormal} and {\rm complete} single-particle basis. This can be verified using the discrete orthogonality property of the Chebyshev polynomials.

Let us prove that $H_{0}$ has an inverse-square form in the original fermion basis. It is obvious that
\begin{eqnarray}
H_{0} &=& \sum^{L-1}_{q=1} \sum^{L}_{j,k} \frac{6}{L} q^{2} \cos(q\theta_{j}) \cos(q\theta_{k}) c^{\dag}_{j,\sigma}c_{k,\sigma} \nonumber \\
&=& \sum^{L}_{j=1} \sum^{L-1}_{q=1} \frac{3}{L} q^{2} \left[ 1+\cos(2q\theta_{j}) \right] c^{\dag}_{j,\sigma}c_{j,\sigma} + \sum^{L}_{j,k=1;j{\neq}k} \sum^{L-1}_{q=1} \frac{3}{L} q^{2} \left[ \cos(q\theta_{j}-q\theta_{k}) + \cos(q\theta_{j}+q\theta_{k}) \right] c^{\dag}_{j,\sigma}c_{k,\sigma} \nonumber \\
&=& \sum^{L}_{j=1} \sum^{L-1}_{q=1} \frac{3}{L} q^{2} \left\{ 1+\cos\left[\frac{q\pi}{L}(2j-1)\right] \right\} c^{\dag}_{j,\sigma}c_{j,\sigma} \nonumber \\
&& + \sum^{L}_{j,k=1;j{\neq}k} \sum^{L-1}_{q=1} \frac{3}{L} q^{2} \left\{ \cos\left[\frac{q\pi}{L}(j-k)\right] + \cos\left[\frac{q\pi}{L}(j+k+1)\right] \right\} c^{\dag}_{j,\sigma}c_{k,\sigma} .
\end{eqnarray}
For an integer $X\in[0,2L)$, we have the summation identity
\begin{eqnarray}
\sum^{L-1}_{q=1} q^{2} \cos\left(\frac{q\pi}{L}X\right) =
\begin{cases}
\frac{(-1)^{X}}{2} L \left[ \frac{1}{\sin^{2}\left(\frac{{\pi}X}{2L}\right)} -1 \right] \quad X{\neq}0 \\
\frac{1}{6} L(L-1)(2L-1) \qquad X=0
\end{cases}.
\end{eqnarray}
This helps us to show that
\begin{eqnarray}
H_{0} &=& \sum^{L}_{j,k=1;j{\neq}k} (-1)^{j-k} \left( \frac{6}{|z_{j}-z_{k}|^2} - \frac{6}{|z_{j}-z^{*}_{k}|^2} \right) c^{\dag}_{j,\sigma}c_{k,\sigma} \nonumber \\
&& + \sum^{L}_{j=1} \left[ L^{2}+\frac{1}{2} - \frac{3}{2\sin^{2}\theta_{j}} \right] c^{\dag}_{j,\sigma}c_{j,\sigma},
\end{eqnarray}
where $z_{j}=\exp(i\theta_{j})$ and $z^{*}_{j}=\exp(-i\theta_{j})$ are the complex coordinates of the site $j$ and its mirror image, respectively.

To compute the results of acting $H_{0}$ on $|\Psi\rangle$, we need to know commutators of the $[H_{0},\zeta^{\dag}_{m}]$ type. This requires linear transformations between the two sets of single-particle orbitals defined by (the spin index $\sigma$ is suppressed below for simplicity)
\begin{eqnarray}
\zeta^{\dag}_{m} = \sum^{L-1}_{q=0} W_{mq} d^{\dag}_{q} \qquad d^{\dag}_{q}=\sum^{L-1}_{m=0} (W^{-1})_{qm} \zeta^{\dag}_{m},
\end{eqnarray}
One can see that
\begin{eqnarray*}
\left[ H_{0},\zeta^{\dag}_{m} \right] &=& \sum^{L-1}_{q,q'=0} 3q^{2} \; W_{mq'} \left[ d^{\dag}_{q} d_{q} , d^{\dag}_{q'} \right]
= \sum^{L-1}_{q=0} 3q^{2} \; W_{mq} d^{\dag}_{q} \\
&=& \sum^{L-1}_{q=0} \sum^{L-1}_{m'=0} 3q^{2} \; W_{mq} \left( W^{-1} \right)_{qm'} \zeta^{\dag}_{m'} = \sum^{L-1}_{m'=0} A_{mm'}\zeta^{\dag}_{m'}
\end{eqnarray*}
with
\begin{eqnarray}
A_{mm'} = \sum^{L-1}_{q=0} 3q^{2} \; W_{mq} \left( W^{-1} \right)_{qm'}.
\end{eqnarray}
For the $m=0$ case, we have $\left[H_{0},\zeta^{\dag}_{0}\right]=0$ because $d^{\dag}_{0}$ is a zero mode of $H_{0}$ and $\zeta^{\dag}_{0}=\sqrt{L}d^{\dag}_{0}$. This helps us to prove that $H_{0}|\Psi\rangle=|\Psi^{\uparrow}_{0}\rangle +|\Psi^{\downarrow}_{0}\rangle$ with
\begin{eqnarray}
|\Psi^{\uparrow}_{0}\rangle &=& c^{\dag}_{0,\uparrow} H_{0} \zeta^{\dag}_{0,\downarrow} |{\rm PFS}\rangle = c^{\dag}_{0,\uparrow} \zeta^{\dag}_{0,\downarrow} H_{0} |{\rm PFS}\rangle \notag \\
&=& c^{\dag}_{0,\uparrow} \left\{ -\zeta^{\dag}_{0,\uparrow}\zeta^{\dag}_{0,\downarrow} \sum^{M-1}_{m=1} \left[ \prod^{m-1}_{\alpha=1} \zeta^{\dag}_{\alpha,\uparrow} \zeta^{\dag}_{\alpha,\downarrow} \right] A_{m0} \zeta^{\dag}_{m,\downarrow} \left[ \prod^{M-1}_{\alpha=m+1} \zeta^{\dag}_{\alpha,\uparrow} \zeta^{\dag}_{\alpha,\downarrow} \right] |0\rangle + 2\zeta^{\dag}_{0,\downarrow} \sum^{M-1}_{m=1} A_{mm}|{\rm PFS}\rangle \right\},
\label{eq:ActionH0}
\end{eqnarray}
so the diagonal entries $A_{mm}$ and the first column $A_{m0}$ of the matrix $A_{mm'}$ are sufficient for our purpose.

Let us consider the transformation matrix from $d^{\dag}_{q}$ to $\zeta^{\dag}_{m}$. The $m=0$ case is simple since $\zeta^{\dag}_{0}=\sum^{L}_{j=1}c^{\dag}_{j}=\sqrt{L}d^{\dag}_{0}$. For $m{\geq}1$, we have
\begin{eqnarray}
\zeta^{\dag}_{m} &=& \sum^{L}_{j=1} \cos^{m-1}\theta_{j} \left( 1-\cos\theta_{j} \right) c^{\dag}_{j} \notag \\
&=& \sum^{L}_{j=1} \left[ \left( \frac{e^{i\theta_{j}}+e^{-i\theta_{j}}}{2} \right)^{m-1} - \left( \frac{e^{i\theta_{j}}+e^{-i\theta_{j}}}{2} \right)^{m}\right] c^{\dag}_{j} \notag \\
&=& \sum_{j=1}^{L}\frac{c^{\dag}_{j}}{2^{m}}\times  \notag \\
&\phantom{=}&
\begin{cases}
2\sum^{\frac{m-1}{2}}_{\alpha=-\frac{m-1}{2}} \dbinom{m-1}{\frac{m-1}{2}+\alpha}e^{i[(\frac{m-1}{2}+\alpha)-(\frac{m-1}{2}-\alpha)]\theta_{j}}-\sum_{\alpha=-\frac{m-1}{2}}^{\frac{m+1}{2}}\dbinom{m}{\frac{m-1}{2}+\alpha}e^{i[(\frac{m-1}{2}+\alpha)-(\frac{m+1}{2}-\alpha)]\theta_{j}} \; \text{odd} \; m{\geq}1 \\
2\sum^{\frac{m}{2}}_{\alpha=-\frac{m}{2}+1} \dbinom{m-1}{\frac{m}{2}-1+\alpha}e^{i[(\frac{m}{2}-1+\alpha)-(\frac{m}{2}-\alpha)]\theta_{j}}-\sum_{\alpha =-\frac{m}{2}}^{\frac{m}{2}}\dbinom{m}{\frac{m}{2}+\alpha} e^{i[(\frac{m}{2}+\alpha)-(\frac{m}{2}-\alpha )]\theta_{j}} \; \text{even} \; m{\geq}2
\end{cases} \notag \\
&=& \sum^{L}_{j=1} \frac{c^{\dag}_{j}}{2^{m}}
\begin{cases}
2\dbinom{m-1}{\frac{m-1}{2}}+4\sum^{\frac{m-1}{2}}_{\alpha=1} \dbinom{m-1}{\frac{m-1}{2}+\alpha} \cos(2\alpha \theta _{j})-2\sum^{\frac{m+1}{2}}_{\alpha=1} \dbinom{m}{\frac{m-1}{2}+\alpha}\cos(2\alpha-1)\theta_{j} \quad \text{odd} \; m{\geq}1 \\
4\sum^{\frac{m}{2}}_{\alpha=1} \dbinom{m-1}{\frac{m}{2}-1+\alpha}\cos(2\alpha-1)\theta_{j}-\dbinom{m}{\frac{m}{2}}-2\sum^{\frac{m}{2}}_{\alpha=1} \dbinom{m}{\frac{m}{2}+\alpha}\cos(2\alpha\theta_{j}) \quad \text{even} \; m{\geq}2
\end{cases} \notag \\
&=& \sqrt{\frac{L}{2}}\frac{1}{2^{m}}
\begin{cases}
2\sqrt{2}\dbinom{m-1}{\frac{m-1}{2}}d^{\dag}_{0}+4\sum^{\frac{m-1}{2}}_{\alpha=1} \dbinom{m-1}{\frac{m-1}{2}+\alpha}d^{\dag}_{2\alpha }-2\sum^{\frac{m+1}{2}}_{\alpha=1} \dbinom{m}{\frac{m-1}{2}+\alpha}d^{\dag}_{2\alpha-1} \quad \text{odd} \; m{\geq}1 \\
4\sum^{\frac{m}{2}}_{\alpha=1} \dbinom{m-1}{\frac{m}{2}-1+\alpha }d^{\dag}_{2\alpha-1}-\sqrt{2}\dbinom{m}{\frac{m}{2}}d^{\dag}_{0}-2\sum^{\frac{m}{2}}_{\alpha=1} \dbinom{m}{\frac{m}{2}+\alpha}d^{\dag}_{2\alpha} \quad \text{even} \; m{\geq}2
\end{cases}.
\end{eqnarray}
The matrix $W_{mq}$ is
\begin{eqnarray}
\begin{cases}
\sqrt{L}\delta_{q0} \quad m=0 \\
\sqrt{\frac{L}{2}}\frac{1}{2^{m}}\left[ 2\sqrt{2}\dbinom{m-1}{\frac{m-1}{2}}\delta_{q0}+4\sum^{\frac{m-1}{2}}_{\alpha=1} \dbinom{m-1}{\frac{m-1}{2}+\alpha}\delta_{q,2\alpha}-2\sum^{\frac{m+1}{2}}_{\alpha=1} \dbinom{m}{\frac{m-1}{2}+\alpha }\delta _{q,2\alpha-1}\right] \quad \text{odd} \; m{\geq}1 \\
\sqrt{\frac{L}{2}}\frac{1}{2^{m}}\left[ -\sqrt{2}\dbinom{m}{\frac{m}{2}}\delta_{q0}+4\sum^{\frac{m}{2}}_{\alpha=1} \dbinom{m-1}{\frac{m}{2}-1+\alpha}\delta_{q,2\alpha-1}-2\sum^{\frac{m}{2}}_{\alpha=1} \dbinom{m}{\frac{m}{2}+\alpha}\delta_{q,2\alpha}\right] \quad \text{even} \; m{\geq}2
\end{cases},
\end{eqnarray}
which has a lower triangular form (i.e., $W_{mq}=0$ if $q>m$)
\begin{eqnarray}
W=\sqrt{\frac{L}{2}}
\begin{pmatrix}
\sqrt{2} &  &  &  &  &  &  &  \\
\sqrt{2} & -1 &  &  &  &  &  &  \\
-\frac{1}{\sqrt{2}} & 1 & -\frac{1}{2} &  &  &  &  &  \\
\frac{1}{\sqrt{2}} & -\frac{3}{4} & \frac{1}{2} & -\frac{1}{4} &  &  &  &
\\
-\frac{3}{4\sqrt{2}} & \frac{3}{4} & -\frac{1}{2} & \frac{1}{4} & -\frac{1}{8%
} &  &  &  \\
\frac{3}{4\sqrt{2}} & -\frac{5}{8} & \frac{1}{2} & -\frac{5}{16} & \frac{1}{8%
} & -\frac{1}{16} &  &  \\
-\frac{5}{8\sqrt{2}} & \frac{5}{8} & -\frac{15}{32} & \frac{5}{16} & -\frac{3%
}{16} & \frac{1}{16} & -\frac{1}{32} &  \\
&  &  & \cdots &  &  &  & \ddots
\end{pmatrix}.
\end{eqnarray}
In particular, $W_{mm}=\sqrt{\frac{L}{2}}(\sqrt{2}\delta_{m0}-\frac{1}{2^{m-1}}\delta_{m{\neq}0})$.

Let us now consider the transformation matrix from $\zeta^{\dag}_{m}$ to $d^{\dag}_{q}$. It should also be a triangular matrix with $(W^{-1})_{qm}=0$ if $m>q$. The $q=0$ case is simple since $d^{\dag}_{0}=\sqrt{\frac{1}{L}}\sum^{L}_{j=1}c^{\dag}_{j}=\sqrt{\frac{1}{L}}\zeta^{\dag}_{0}$. For $q>0$, we have
\begin{eqnarray}
d^{\dag}_{q>0} &=& \sqrt{\frac{2}{L}} \sum^{L}_{j=1} \cos(q\theta_{j}) c^{\dag}_{j} = \sqrt{\frac{2}{L}} \frac{q}{2} \sum^{\lfloor{q/2}\rfloor}_{k=0} (-1)^{k} \frac{(q-k-1)!}{k!(q-2k)!}2^{q-2k} \sum^{L}_{j=1} \cos^{q-2k} \theta_{j} c^{\dag}_{j} \\
&=& \sqrt{\frac{2}{L}} \frac{q}{2} \sum^{\lfloor{q/2}\rfloor}_{k=0} (-1)^{k} \frac{(q-k-1)!}{k!(q-2k)!}2^{q-2k} \left(\zeta^{\dag}_{0}-\sum^{q-2k}_{p=1}\zeta^{\dag}_{p}\right)  \notag \\
&=& \sum^{q}_{m=0} (W^{-1})_{q>0,m}\zeta^{\dag}_{m} .
\end{eqnarray}
The second equal sign is true because $\cos(q\theta_{j})=T_{q}(\cos\theta_{j})$, where
\begin{eqnarray}
T_{q}(x)=\frac{q}{2} \sum^{\lfloor{q/2}\rfloor}_{k=0} (-1)^{k}\frac{(q-k-1)!}{k!(q-2k)!}(2x)^{q-2k}
\end{eqnarray}
is the Chebyshev polynomial. The useful matrix elements are
\begin{equation}
(W^{-1})_{q>0,q>0}=-\sqrt{\frac{2}{L}}2^{q-1}
\end{equation}
and
\begin{eqnarray}
(W^{-1})_{q>0,0} &=& \sqrt{\frac{2}{L}}\frac{q}{2} \sum^{\lfloor{q/2}\rfloor}_{k=0} (-1)^{k} \frac{(q-k-1)!}{k!(q-2k)!}2^{q-2k} = \sqrt{\frac{2}{L}}.
\end{eqnarray}

The transformation matrices derived above can be used to prove that
\begin{eqnarray}
A_{m>0,m>0} &=& \sum^{L-1}_{q=0} 3q^{2} \; W_{mq}(W^{-1})_{qm} = 3m^{2} \; W_{m>0,m>0}(W^{-1})_{m>0,m>0} = 3m^{2}
\label{eq:matAdiagonal}
\end{eqnarray}
and
\begin{eqnarray}
A_{m>0,0} &=& \sum^{L-1}_{q=0} 3q^{2} \; W_{mq}(W^{-1})_{q0} = \sum^{L-1}_{q=1} 3q^{2} \; W_{m>0,q}(W^{-1})_{q0} \notag \\
&=& \sum^{L-1}_{q=1} 3q^{2}
\begin{cases}
\frac{1}{2^{m}}\left[ 2\sqrt{2}\dbinom{m-1}{\frac{m-1}{2}}\delta_{q0}+4\sum^{\frac{m-1}{2}}_{\alpha=1} \dbinom{m-1}{\frac{m-1}{2}+\alpha}\delta_{q,2\alpha}-2\sum^{\frac{m+1}{2}}_{\alpha=1} \dbinom{m}{\frac{m-1}{2}+\alpha}\delta_{q,2\alpha -1}\right] \quad \text{odd} \; m{\geq}1 \\
\frac{1}{2^{m}}\left[-\sqrt{2}\dbinom{m}{\frac{m}{2}}\delta_{q0}+4\sum^{\frac{m}{2}}_{\alpha=1} \dbinom{m-1}{\frac{m}{2}-1+\alpha}\delta_{q,2\alpha-1}-2\sum^{\frac{m}{2}}_{\alpha=1} \dbinom{m}{\frac{m}{2}+\alpha}\delta_{q,2\alpha}\right] \quad \text{even} \; m{\geq}2
\end{cases} \notag \\
&=& \frac{1}{2^{m}}
\begin{cases}
\sum^{\frac{m-1}{2}}_{\alpha=1} 48\alpha^{2} \dbinom{m-1}{\frac{m-1}{2}+\alpha}-\sum^{\frac{m+1}{2}}_{\alpha=1} 6(2\alpha-1)^{2}\dbinom{m}{\frac{m-1}{2}+\alpha} \quad \text{odd} \; m{\geq}1 \\
\sum^{\frac{m}{2}}_{\alpha=1} 12(2\alpha-1)^{2} \dbinom{m-1}{\frac{m}{2}-1+\alpha}-\sum^{\frac{m}{2}}_{\alpha=1} 24\alpha^{2}\dbinom{m}{\frac{m}{2}+\alpha} \quad \text{even} \; m{\geq}2
\end{cases} \notag\\
&=& \frac{1}{2^{m}}
\begin{cases}
6(m-1)2^{m-1}-6m2^{m-1} \\
12(m-1)2^{m-2}-3m2^{m}%
\end{cases} \notag \\
&=& -3 ,
\label{eq:matAcolumn}
\end{eqnarray}
where certain summation identities from Sec.~B4 are used. For comparison, the matrix $A$ computed using \textit{Mathematica} is
\begin{eqnarray}
A=
\begin{pmatrix}
0 &  &  &  &  &  &  &  \\
-3 & 3 &  &  &  &  &  &  \\
-3 & 9 & 12 &  &  &  &  &  \\
-3 & -3 & 15 & 27 &  &  &  &  \\
-3 & 3 & -15 & 21 & 48 &  &  &  \\
-3 & 3 & 3 & -33 & 27 & 75 &  &  \\
-3 & 3 & 3 & 3 & -57 & 33 & 108 &  \\
&  &  & \cdots &  &  &  & \ddots%
\end{pmatrix} .
\end{eqnarray}

Finally, we arrive at
\begin{eqnarray}
|\Psi^{\uparrow}_{0}\rangle &=& c^{\dag}_{0,\uparrow} \left\{ -\zeta^{\dag}_{0,\uparrow}\zeta^{\dag}_{0,\downarrow} \sum^{M-1}_{m=1} \left[ \prod^{m-1}_{\alpha=1} \zeta^{\dag}_{\alpha,\uparrow} \zeta^{\dag}_{\alpha,\downarrow} \right] A_{m0} \zeta^{\dag}_{m,\downarrow} \left[ \prod^{M-1}_{\alpha=m+1} \zeta^{\dag}_{\alpha,\uparrow} \zeta^{\dag}_{\alpha,\downarrow} \right] |0\rangle + 2\zeta^{\dag}_{0,\downarrow} \sum^{M-1}_{m=1} A_{mm}|{\rm PFS}\rangle \right\} \notag \\
&=& c^{\dag}_{0,\uparrow} \left\{ 3\zeta^{\dag}_{0,\uparrow}\zeta^{\dag}_{0,\downarrow} \sum^{M-1}_{m=1} \left[ \prod^{m-1}_{\alpha=1} \zeta^{\dag}_{\alpha,\uparrow} \zeta^{\dag}_{\alpha,\downarrow} \right] \zeta^{\dag}_{m,\downarrow} \left[ \prod^{M-1}_{\alpha=m+1} \zeta^{\dag}_{\alpha,\uparrow} \zeta^{\dag}_{\alpha,\downarrow} \right] |0\rangle + \zeta^{\dag}_{0,\downarrow} \sum^{M-1}_{m=1} 6m^{2} |{\rm PFS}\rangle \right\}.
\end{eqnarray}
Its first part and the unwanted particle-hole excitations in Eq.~(\ref{eq:kondo-chemical-unwanted}) cancel each other. It is then obvious that
\begin{eqnarray}
\left( H_{0}+H_{\rm P}+H_{\rm K} \right) |\Psi\rangle = \left[ \sum^{M-1}_{m=1} 6m^{2} - \frac{3}{2}(M-1) \right] \left( c^{\dag}_{0,\uparrow}\zeta^{\dag}_{0,\downarrow} - c^{\dag}_{0,\downarrow}\zeta^{\dag}_{0,\uparrow} \right) |{\rm PFS}\rangle ,
\end{eqnarray}
so the target state $|\Psi\rangle$ is indeed an eigenstate with eigenvalue
\begin{eqnarray}
E(M) = \sum^{M-1}_{m=1} 6m^{2}-\frac{3}{2}(M-1) = (M-1) \left( 2M^{2}-M-\frac{3}{2}\right).
\end{eqnarray}

\subsection{B4: Some useful identities}

This subsection computes the sums in (\ref{eq:matAcolumn}). For odd $m{\geq}1$, we introduce an auxiliary function
\begin{eqnarray}
(x+x^{-1})^{m-1} &=& \sum^{\frac{m-1}{2}}_{\alpha=-\frac{m-1}{2}} \dbinom{m-1}{\frac{m-1}{2}+\alpha }x^{\frac{m-1}{2}+\alpha}x^{-(\frac{m-1}{2}-\alpha)} \notag \\
&=& \sum^{\frac{m-1}{2}}_{\alpha=-\frac{m-1}{2}} \dbinom{m-1}{\frac{m-1}{2}+\alpha}x^{2\alpha} = \dbinom{m-1}{\frac{m-1}{2}}+\sum^{\frac{m-1}{2}}_{\alpha=1} \dbinom{m-1}{\frac{m-1}{2}+\alpha}(x^{2\alpha}+x^{-2\alpha}).
\end{eqnarray}
Its derivatives are
\begin{eqnarray}
(m-1)(x+x^{-1})^{m-2}(1-x^{-2})=\sum^{\frac{m-1}{2}}_{\alpha=1} \dbinom{m-1}{\frac{m-1}{2}+\alpha} 2\alpha \left( x^{2\alpha -1}-x^{-2\alpha-1} \right)
\end{eqnarray}
and
\begin{eqnarray}
&& (m-1)(m-2)(x+x^{-1})^{m-3}(1-x^{-2})^{2}+(m-1)(x+x^{-1})^{m-2}\frac{2}{x^{3}} \notag \\
= && \sum^{\frac{m-1}{2}}_{\alpha=1} \dbinom{m-1}{\frac{m-1}{2}+\alpha} 2\alpha \left[ (2\alpha-1)x^{2\alpha -2}+(2\alpha+1)x^{-2\alpha-2} \right].
\label{eq:sumidentity1}
\end{eqnarray}
If $x$ is chosen to be $1$ in Eq.~(\ref{eq:sumidentity1}), we obtain
\begin{eqnarray}
(m-1)2^{m-1}=\sum^{\frac{m-1}{2}}_{\alpha=1} \dbinom{m-1}{\frac{m-1}{2}+\alpha}8\alpha^{2}.
\label{eq:sumidentity2}
\end{eqnarray}

For odd $m$, we study
\begin{eqnarray*}
(x+x^{-1})^{m} &=& \sum^{\frac{m+1}{2}}_{\alpha=-\frac{m-1}{2}} \dbinom{m}{\frac{m-1}{2}+\alpha }x^{\frac{m-1}{2}+\alpha }x^{-(\frac{m+1}{2}-\alpha)} \\
&=& \sum^{\frac{m+1}{2}}_{\alpha=-\frac{m-1}{2}} \dbinom{m}{\frac{m-1}{2}+\alpha}x^{2\alpha-1} = \sum^{\frac{m+1}{2}}_{\alpha=1} \dbinom{m}{\frac{m-1}{2}+\alpha}(x^{2\alpha-1}+x^{-2\alpha+1}) .
\end{eqnarray*}
Its derivatives are
\begin{eqnarray}
m(x+x^{-1})^{m-1}(1-x^{-2}) = \sum^{\frac{m+1}{2}}_{\alpha=1} \dbinom{m}{\frac{m-1}{2}+\alpha} \left[ (2\alpha-1)x^{2\alpha-2}-(2\alpha-1)x^{-2\alpha} \right],
\end{eqnarray}
and
\begin{eqnarray}
&& m(m-1)(x+x^{-1})^{m-2}(1-x^{-2})^{2}+m(x+x^{-1})^{m-1}\frac{2}{x^{3}} \notag \\
= && \sum^{\frac{m+1}{2}}_{\alpha=1} \dbinom{m}{\frac{m-1}{2}+\alpha} \left[ (2\alpha-1)(2\alpha-2)x^{2\alpha-3}+(2\alpha-1)2\alpha x^{-2\alpha-1} \right].
\label{eq:sumidentity3}
\end{eqnarray}
If $x$ is chosen to be $1$ in Eq.~(\ref{eq:sumidentity3}), we obtain
\begin{eqnarray}
m2^{m-1}=\sum^{\frac{m+1}{2}}_{\alpha=1} \dbinom{m}{\frac{m-1}{2}+\alpha}(2\alpha-1)^{2}.
\label{eq:sumidentity4}
\end{eqnarray}

The sums with even $m$ can be obtained from previous ones. In Eq.~(\ref{eq:sumidentity2}), replacing $m-1$ with $m$ yields
\begin{eqnarray}
m2^{m} = \sum^{\frac{m}{2}}_{\alpha=1} \dbinom{m}{\frac{m}{2}+\alpha}8\alpha^{2} .
\label{eq:sumidentity5}
\end{eqnarray}
In Eq.~(\ref{eq:sumidentity4}), replacing $m+1$ with $m$ yields
\begin{equation}
(m-1)2^{m-2} = \sum^{\frac{m}{2}}_{\alpha=1} \dbinom{m-1}{\frac{m}{2}-1+\alpha}(2\alpha-1)^{2}.
\label{eq:sumidentity6}
\end{equation}

\section{Appendix C: Spin-Spin Correlation Function}

This section provides more details about how to compute the spin-spin correlation function of the ground state. To begin with, we compute the overlap of two generic free fermion states. For a set of fermionic operators $c^{\dag}_{j}$ and $c_{j}$ satisfying the anticommutation relations, we define two sets of operators $\alpha^{\dag}_{a}$ and $\beta^{\dag}_{b}$
\begin{eqnarray}
\alpha_{a} = \sum_{j} U_{aj} c_{j}, \quad \beta_{b} = \sum_{j} V_{bj} c_{j},
\label{eq:ss1}
\end{eqnarray}
where $U_{aj}$ and $V_{bj}$ are two matrices that may not be isometry. It is easy to see that
\begin{eqnarray}
\langle 0 | \alpha_{a} \beta^{\dag}_{b} |0 \rangle = \sum_{jk} U_{aj} V^{*}_{bk} \langle{0}| c_{j} c^{\dag}_{k} |{0}\rangle = \sum_{j} U_{aj} V^{*}_{bj} = \sum_{j} \left[ UV^{\dag} \right]_{ab} = G_{ab}
\label{eq:ss2}
\end{eqnarray}
with $G=UV^{\dag}$. The overlap between $\alpha^{\dag}_{1} \alpha^{\dag}_{2} \cdots \alpha^{\dag}_{M} |0\rangle$ and $\beta^{\dag}_{1} \beta^{\dag}_{2} \cdots \beta^{\dag}_{M} |0\rangle$ is found to be
\begin{eqnarray}
&& \langle 0 | \alpha_{M} \cdots \alpha_{2}\alpha_{1} \beta^{\dag}_{1} \beta^{\dag}_{2} \cdots \beta_{M} |0 \rangle \nonumber \\
&=& \langle{0}| \alpha_{1} \beta^{\dag}_{1} |{0}\rangle \langle{0}| \alpha_{2} \beta^{\dag}_{2} |{0}\rangle \cdots \langle{0}| \alpha_{M} \beta^{\dag}_{M} |{0}\rangle + \text{all possible permutations with signs} \nonumber \\
&=& G_{11} G_{22} \cdots G_{MM} + \text{all possible permutations with signs} \nonumber \\
&=& \det G
\label{eq:ss3}
\end{eqnarray}
using Eq.~(\ref{eq:ss2}) and Wick's theorem.

The numerator of the spin-spin correlation function can be simplified using the SU(2) spin symmetry as
\begin{eqnarray}
\langle\Psi| {\bf S}_{0}\cdot{\bf S}_{j} |\Psi\rangle = \langle\Psi| {\bf S}_{0}\cdot{\bf S}_{j} |\Psi\rangle = 3 \langle\Psi| S^{x}_{0}S^{x}_{j} |\Psi\rangle = \frac{3}{2} \left( \langle\Psi| S^{x}_{0}S^{x}_{j} |\Psi\rangle + \langle\Psi| S^{y}_{0}S^{y}_{j} |\Psi\rangle \right) = \frac{3}{2} \langle\Psi| S^{+}_{0}S^{-}_{j} |\Psi\rangle
\end{eqnarray}
with
\begin{eqnarray}
\langle\Psi| S^{+}_{0}S^{-}_{j} |\Psi\rangle &=& \langle{0}| \prod'^{1}_{m=M-1} \prod_{\sigma=\downarrow,\uparrow} \zeta_{m,\sigma} \zeta_{0,\uparrow} \; c_{j,\uparrow} c^{\dag}_{j,\downarrow} \; \zeta^{\dag}_{0,\uparrow} \prod^{M-1}_{m=1} \prod_{\sigma=\uparrow,\downarrow} \zeta^{\dag}_{m,\sigma} |{0}\rangle \notag \\
&=& - \langle{0}| \prod'^{1}_{m=M-1} \zeta_{m,\uparrow} \; c_{j,\uparrow} \zeta^{\dag}_{0,\uparrow} \; \prod^{M-1}_{m=1} \zeta^{\dag}_{m,\uparrow} |{0}\rangle \cdot \langle{0}| \prod'^{1}_{m=M-1} \zeta_{m,\downarrow} \zeta_{0,\downarrow} \; c^{\dag}_{j,\downarrow} \; \prod^{M-1}_{m=1} \zeta^{\dag}_{m,\downarrow} |{0}\rangle ,
\end{eqnarray}
where $\prod'^{1}_{m=M-1} \prod_{\sigma=\downarrow,\uparrow} \zeta_{m,\sigma}$ denotes reversed order of products. The denominator of the spin-spin correlation function is the unnormalized overlap
\begin{eqnarray}
\langle\Psi|\Psi\rangle &=& \langle{0}| \prod'^{1}_{m=M-1} \zeta_{m,\uparrow} \; \prod^{M-1}_{m=1} \zeta^{\dag}_{m,\uparrow} |{0}\rangle \cdot \langle{0}| \prod'^{1}_{m=M-1} \zeta_{m,\downarrow} \; \zeta_{0,\downarrow} \zeta^{\dag}_{0,\downarrow} \; \prod^{M-1}_{m=1} \zeta^{\dag}_{m,\downarrow} |{0}\rangle \notag \\
&\phantom{=}& + \langle{0}| \prod'^{1}_{m=M-1} \zeta_{m,\uparrow} \; \zeta_{0,\uparrow} \zeta^{\dag}_{0,\uparrow} \; \prod^{M-1}_{m=1} \zeta^{\dag}_{m,\uparrow} |{0}\rangle \cdot \langle{0}| \prod'^{1}_{m=M-1} \zeta_{m,\downarrow} \; \prod^{M-1}_{m=1} \zeta^{\dag}_{m,\downarrow} |{0}\rangle .
\end{eqnarray}
Both of them can be computed using Eq.~(\ref{eq:ss3}).

\end{widetext}

\end{document}